 \documentstyle[12pt,twoside,fleqn,espcrc1]{article}

\newcommand{\AmS}{{\protect\the\textfont2
  A\kern-.1667em\lower.5ex\hbox{M}\kern-.125emS}}

\title{Role of relativity and nucleon compositeness in few-body systems}

\author{S. J. Wallace 
\address{Department of Physics and Center for Theoretical Physics\\
University of Maryland, College Park MD 20742-4111 }%
       \thanks{Support of the U.S. Dept. of Energy under grant 
DE-FG02-93ER-40762 is gratefully acknowledged.}}
       
\begin{document}

\maketitle

\begin{abstract}
Recent progress is reviewed in four areas where new experimental data
have been obtained: proton-proton bremsstrahlung, threshold pion production
in proton-proton collisions, elastic electron-deuteron scattering and 
deuteron photodisintegration at several GeV photon energy.  
Relativistic effects are expected to be significant in these processes.
High energy photodisintegration suggests that partonic behavior
could be relevant in subamplitudes at large momentum transfer. 
\end{abstract}

\section{INTRODUCTION}

This review of recent work in
the field of few-body systems follows an earlier review\cite{Wallace97} in focusing on
particular experiments whose interpretations are expected to involve 
relativistic effects along with other physics. 
Relativistic effects generally are expected to be significant in the
dynamics of few-body systems when one or more momenta involved are comparable
to the nucleon's mass.  Even in cases where a nonrelativistic interpretation
is successful, a relativistic treatment is preferred provided that
it contains all the relevant physics.  These points motivate much work in 
the field, but it must be recognized that identifying a relativistic effect 
can be a subtle matter.  

Depending upon the form and organization of the dynamics, relativistic
effects may take different forms.  For example, there are several ways
to reduce the four-dimension Bethe-Salpeter equation to three
dimensions.\cite{Logunov63,Blankenbecler66,Gross69.82,Phillips96}  
Matrix elements corresponding to physical observables
should be the same independent of which reduction is used provided
that equivalent lagrangians are used and all relevant terms are kept.
Effects in one form of dynamics, such as Z-graphs or off-shell effects
in a subamplitude or short-range meson-exchange currents, 
may be replaced by contact terms in another form of
dynamics, with similar results.  Effects of boosts may be in the wave
functions in one analysis and in the currents in another, again with
similar results for matrix elements.  It is essential to understand these
differences in order to see how pieces of the physics get shuffled from one
place to another by the choice of formalism.   Equivalent results
should be found if each formalism is fully evaluated.    
Relativistic effects can be different 
depending on the formalism used.    

Although different forms of dynamics applied to the same lagrangian
should produce equivalent results, this fails if currents are not 
consistent with the interactions between the particles. 
One of the main points of relativistic formulations based on
a meson-exchange lagrangian is to be able to apply Mandelstam's 
construction of currents\cite{Mandelstam55}
 for the Bethe-Salpeter equation\cite{Salpeter51}
in order to formulate currents that are consistent with 
interactions.  It becomes interesting to discuss the physics 
differences between various analyses once this has been done,
the goal being to discover a lagrangian that works to describe the 
physics independently of the formalism used to evaluate it.

In recent years, considerable attention has been given to the
problem of formulating consistent currents.
Arenh\"{o}vel and collaborators have developed consistent
currents for the analyses of electromagnetic interactions of the
two-nucleon system by considering a rather 
complete set of $1/M$ corrections
to the nonrelativistic formalism.\cite{Ritz97} 
Reductions of the Bethe-Salpeter
equation to three-dimensions also require a reduction of the currents
to three dimensions.  
Gross and Riska\cite{GrossRiska87}, and more 
recently Adam, Van Orden and Gross\cite{Adam97a}, and 
Blankleider and Kvinkhidze\cite{Kvinikhidze98} have 
discussed the appropriate current for the 
spectator formalism, in which one particle is on its mass shell. 
The spectator formalism requires a careful consideration of the
case where the photon is absorbed by the on-shell nucleon.
Phillips and Wallace\cite{PW98} have developed 
the appropriate currents for the equal-time formalism.
An important result in Refs.\cite{Adam97a,PW98} 
is that truncating the interaction 
at a definite order in the coupling constants, 
and truncating the exchange currents at the same order,
preserves the Ward-Takahashi identities.
This ensures that the formalism has conserved 
currents and is well suited to
phenomenology based on the one-boson-exchange truncation.  Works by 
Carbonell, Desplanques, Karmanov and Mathiot\cite{Carbonell98} and 
de Melo, Frederico, Naus and Sauer\cite{Sauer99} have discussed
the currents for light-front dynamics.
These authors find that contact terms are required in the light-front
formalism in order to replace Z-graphs of a covariant formalism.
This is an example of a relativistic 
effect that takes a different form in different dynamics.

\section{BREMSSTRAHLUNG}

Recent experiments have provided precise bremsstrahlung data for 
proton-proton collisions using a 190 MeV polarized proton 
beam.\cite{KVIbremss1,KVIbremss2}  Data at 389 MeV
also have been obtained at RCNP Osaka\cite{RCNPbremss}
and an experiment at 300 MeV\cite{Bilger98} has been performed at COSY in order
to check older TRIUMF measurements.\cite{Michaelian90}
At the same time, questions have been raised
about the motivations for studying the bremsstrahlung 
process.\cite{FearingPRL98,FearingFBS99,FearingScherer00} 

Historically, bremsstrahlung experiments have been motivated by a
desire to access off-shell matrix elements of the $NN$ interactions,
with the expectation that this would help to discriminate between
potential models that yield equivalent descriptions of the $NN$ data.
Fearing\cite{FearingPRL98,FearingFBS99} has pointed 
out that this motivation is flawed
because off-shell effects are ambiguous.  When one starts from a
lagrangian, field transformations may be used to shuffle off-shell
effects in one analysis into contact terms in another analysis that is
based on an equivalent lagrangian.\cite{FearingScherer00}  
Equivalence means that all
S-matrix elements are the same for the two lagrangians.  
Off-shell effects are not the same for equivalent lagrangians, 
which makes them ambiguous and unmeasurable.

The arguments of Fearing and Scherer do not take into account a
standard convention that constrains the ambiguity in nucleon-nucleon
bremsstrahlung.  Generally, models of the $NN$ interaction are based
on an integral equation with a two-nucleon Green's function, $G$, and
the Green's function is constrained to have a standard form when both
nucleons are in positive-energy states.  In order to maintain the 
standard form for $G$, interactions have to absorb the
effects of field transformations.  Otherwise, one would add
nonstandard terms to $G$. 
Likewise, the currents for nucleons in
positive-energy states are constrained to be consistent with the
standard from of $G$, which requires the effects of field transformations 
to be absorbed into exchange currents
and pair currents.  Nevertheless, significant ambiguities remain.

One well-known source of ambiguity is the fact that there are
unitary equivalences between interactions that reproduce
the same $NN$ data.  
A second source of ambiguity arises because currents cannot
be formulated consistently for phenomenological potential models
because one does not have an underlying lagrangian.  It is possible
to prescribe a connection between the various spin- and isospin-dependent terms in
the phenomenological $NN$ potential and the usual set of meson
exchanges.  Applying the Mandelstam procedure to these parts of the
potential as if they were caused by meson exchanges provides
a prescription for construction of currrents.\cite{Riska89}   
However, there is no clear way to avoid ambiguities
in such a prescription.  
Because of the ambiguities, the motivation for studying bremsstrahlung
has to be to find a lagrangian that works, as discussed in the
Introduction. 

Low's soft-photon theorem \cite{Low58} shows that bremsstrahlung has
large contributions at low photon momenta that are determined by the
on-shell $NN$ scattering amplitudes and model-independent factors
involving $k^{-1}$ and $k^0$, where $k$ is the photon momentum.  The
recent KVI experiments use a 190 MeV proton beam and detect
the two final-state protons, allowing a reconstruction of the  
bremsstrahlung photon.  A photon energy of about 60 MeV is 
typical and this is expected to provide sensitivity to effects beyond
those controlled by the low-energy theorem.  Experimental results have
been compared with theoretical calculations of 
Martinus et al.\cite{Martinus97,Martinus98} that are based on a
relativistic meson-exchange model, 
including important contributions from $\Delta$ currents.
The theoretical
results of Martinus et al. differ significantly from the data at 
forward angles, but agree better at larger angles and for 
the analyzing power.
The forward angle discrepancy seems to be
because the $NN$ model used\cite{FleischerTjon} does not provide
accurate phase shifts at the relevant energies.  Improvement of the
description of phase shifts yields improved results for 
bremsstrahlung.\cite{TjonPC}
The experimental cross section data are in very good agreement with an
analysis based on a soft-photon approximation.\cite{Liou96.95,Korchin96}  However,
the soft-photon approximation does not describe the analyzing power well.
The soft-photon approximation 
takes into account terms beyond those controlled by Low's theorem
because on-shell $NN$ amplitudes are used at energies and angles
appropriate to both the initial and final states of the $NN$ system in
the bremsstrahlung process.

The impressive precision of the new bremsstrahlung data from KVI makes it
possible to 
detect interesting effects of relativistic dynamics,
$\Delta$ currents and meson-exchange currents.  
Previous calculations of these 
effects\cite{Martinus97,deJong96,Eden95.96} have shown 
them to be large compared with what can be
resolved by the new data. 
It remains to be seen whether improved theoretical calculations
will provide a 
good understanding of the new bremsstrahlung experiments.

\section{THRESHOLD PRODUCTION OF NEUTRAL PIONS}

It is now ten years since precise experiments on threshold production
of $\pi^0$ mesons were performed using a cooled proton beam and a
hydrogen target at the Indiana University Cyclotron
Facility.\cite{Meyer90,Meyer92}  The surprise was that cross
sections for $p p \rightarrow p p \pi^0$ were 
five times larger than existing theoretical
predictions.\cite{Koltun66,Miller91}  
A confirmation of the Indiana experiment
was obtained at CELSIUS.\cite{Bondar95}

Over the past ten years, there has been substantial theoretical
interest in the $\pi^0$ production process.  Theoretical 
calculations all have shown that the best understood mechanisms for soft pion
production are suppressed in the reaction.  This was confirmed by several
groups using chiral perturbation theory\cite{Cohen96,Park96,Sato97,Meissner98}
and is also found in
phenomenological models.\cite{Hanhart95,Hanhart98,Hanhart00a}  
Soft-pion contributions are small in part
because of cancellations between the amplitude for direct pion
emission and the rescattering amplitude in which the pion is emitted
by one nucleon and scatters from the other.  An important point is
that the proton momentum required for pion production is at least 
$\sqrt {m_{\pi}M} \approx 2.6 m_\pi$,
which is not small in comparison with the pion 
mass.\cite{Cohen96,Meissner98}  
 A recent
analysis\cite{Dmitrasinovic99} indicates that chiral perturbation theory
converges poorly for the reaction.  Because the
soft pion contributions are much too small to explain the data,
a new term must be added to the chiral lagrangian to sum up all effects of
short-range pion production.  This term is fit to the experimental data.
\cite{vanKolck96}

As first pointed out by Lee and Riska\cite{Lee93},
suppression of effects from soft pions allows smaller effects from
short-range physics to become prominent.  
Lee and Riska showed that the data could be explained by
including a short-range contribution to the axial charge operator,
namely, pion production from an intermediate $N\overline{N}$ pair state
that is produced by exchange of $\sigma$ and $\omega$ mesons.
The importance of this short-range exchange current was confirmed in 
calculations by Horowitz and collaborators\cite{Horowitz94}.  However, both 
analyses used a perturbative treatment of the relativistic effects 
from Z-graphs.
The validity of the perturbative approach to including short-range effects
such as Z-graphs was questioned by Adam et al.\cite{Adam97b} using the spectator 
formalism of Gross.  

Various refinements and alternative mechanisms for threshold pion production
have been proposed, such as couplings to the $N\Delta$ channels as intermediate
states. This is motivated by a recoil contribution in the 
$NN \rightarrow N\Delta$ transition that can be significant 
near threshold\cite{Hernandez95}.  
Using a model of the $\pi N$ scattering amplitude that fits the experimental
data, and includes off-shell effects, a Julich collaboration
\cite{Hanhart95} reinvestigated
the pionic mechanisms and $\Delta$-recoil contributions with the
conclusion that the $\Delta$-recoil contribution is not large enough to 
explain the cross section data.  

The major progress in the past few years has been the measurement of
proton-spin observables in $\pi^0$ production near threshold
at Indiana.\cite{Meyer98,Meyer99,Saha99}  The experiments were designed to
allow a determination  
the separate contributions of s-waves and p-waves, assuming these
to be the only contributing partial waves.  
The results provide a better understanding 
of the reaction and of the role of the $\Delta$ isobar in it.
The theoretical model of the Julich 
group\cite{Hanhart98,Hanhart00a} has been extended to allow
calculations of spin observables also, however, no other calculations 
of spin observables are available.  

The Julich group includes the
standard pion emission, pion rescattering and pion production mechanisms involving
intermediate $\Delta$ states.  These ingredients produce a theoretical cross section that is
too small by about a factor of two.  Including a short-range contribution
from Z-graphs allows the cross section data to be fit by
adjusting the strength of the short-range contribution.  
Moreover, the spin observables are
also reasonably well described.  The calculations indicate that for
pion momenta below 0.8 $m_\pi$, the $\Delta$ contribution
to cross sections is small in comparison with the needed short-range contribution.

  Data for spin observables show trends as pion momentum increases  
that are similar to trends in the  
calculations of the Julich group stemming from the p-wave 
contributions of the $\Delta$.  
However, the trends start at somewhat
lower pion momentum in the data than in the theoretical 
calculations.  This would be consistent with  
p-wave contributions being important somewhat closer to 
threshold than the theoretical model suggests.
There has been a report of D-wave production of $\pi^0$  
near threshold based upon very precise measurements of angular 
distributions.\cite{Zlomanczuk98}    

The clear understanding that has emerged over the last ten years of work
is that s-wave threshold $\pi^0$ production
requires a short-range contribution that may take one or another
equivalent form, i.e., a pion-production contact term, or
meson-exchange currents involving Z-graphs or other meson-exchange
currents.  An example of other meson-exchange currents is shown in the
work of Riska and collaborators\cite{Pena99}, who 
show that the $\rho \pi \omega$
exchange current can contribute significantly, but intermediate $N^*$
resonances provide a small contribution. Thus, the precise nature of
the short-range contribution is not settled but all analyses show that
there must be one.  The $\pi^0$ production at threshold has provided a
unique window to short-range effects in the $NN$ interaction.

\section{ELASTIC ELECTRON-DEUTERON SCATTERING}

Major experimental progress has been made in electron-deuteron
scattering since the review at the Groningen meeting.
Experiments have been completed at the Thomas Jefferson National Accelerator
Facility that measure the deuteron's $A$ form factor and the tensor
alignment parameter $t_{20}$ for $Q > 1 GeV/c$.  
\cite{JlabHallA-t20,JlabHallA-A(Q),JlabHallC-A(Q)}
New data for $B(Q^2)$ 
are expected to be published soon.  
	
The large momentum transfers involved in recent experiments motivate
relativistic treatments of the deuteron and its electromagnetic interactions.
As noted in Ref. \cite{Wallace97}, measurements of the longitudinal-transverse 
asymmetry, $A_{\phi}$, in (e,e$'$p) 
reactions\cite{Schaar92,Steenhoven94} have provided clear evidence 
that current matrix elements of a nucleon in a deuteron should be treated
within a relativistic formalism that incorporates at least positive-energy
Dirac spinors, or equivalent relativistic effects. The low-energy theorem
of Refs. \cite{Wallace95.96,Phillips97} shows that for a composite nucleon, there are 
contributions to second-order
interactions from contact terms, off-shell effects and the composite-particle
Z-graph.  For scalar and vector interaction at low energy,
the sum of such terms produces the same effect
as is obtained from a Z-graph for an elementary particle.  This suggests
use of the Dirac propagator for a nucleon as 
an efficient means to include the model-independent effects.  However,
it is important to respect chiral symmetry when negative-energy 
components are included by using pseudovector $\pi N$ coupling.

Nonrelativistic analyses that include leading-order relativistic
corrections also provide a consistent analysis for smaller $Q$, i.e.,
$Q \leq M$.  A recent analysis of elastic electron scattering from the
deuteron by Arenh\"{o}vel, Ritz and Wilbois\cite{ARW00} is based on a
meson-exchange model that takes into account a rather complete set of
relativistic and exchange-current corrections.  Essentially the same
model has been used in analyses of electrodisintegration and
photodisintegration of the deuteron.\cite{RAW99.98,Ritz97} 
It provides a particularly consistent
framework for the low $Q$ regime.

Over the past ten years, there have been three relativistic analyses of 
$e + D$ elastic scattering  based on
quantum field theory, assuming an effective $NN$ interaction based on
exchange of mesons.  Hummel and Tjon\cite{Hummel899094} used a 
Blankenbecler and Sugar reduction to three dimensions, Gross, Devine 
and Van Orden\cite{VanOrden95} used the spectator 
formalism with one particle on mass shell,
and Phillips, Wallace and Devine\cite{PWD98,PWD99} have used an equal-time (ET) 
reduction to three dimensions. 
Although somewhat different relativistic 
equations are used in each analysis, they have many common features:
each corresponds to a reduction of
the Bethe-Salpeter equation to three dimensions and each  
incorporates Dirac-spinor wave functions for the interacting nucleons, 
including negative-energy components.  With such wave functions, pair
currents are included automatically when the impulse approximation current is
calculated.  
Mandelstam's construction of the current\cite{Mandelstam55} has been used to derive
consistent currents for the three-dimensional reductions and this makes
them attractive theoretically.

Light-front calculations have been performed for
electron-deuteron scattering recently by Carbonell and 
Kharmanov.\cite{CarbonellKarmanov99} In this 
case it is necessary to include
contact terms in order to reproduce the effects of pair currents.  In
principle, it should be possible to define the consistent currents for the
light-front approach similarly to what has been done for the other
formalisms that are based on quantum field theory.  
An alternative approach
to light-front calculations is provided by Chung et
al.\cite{Chung88}   Whether or not exchange currents can be constructed in this
approach so as to achieve equivalence with the others is an open
question.

The most consistent relativistic calculations based on a 
meson-exchange lagrangian have been
performed by Gross, Devine and Van Orden.\cite{VanOrden95}  
Coupling constants and other parameters 
of the meson-exchange model were fit so as to achieve a good description 
of the modern $NN$ phase shifts.\cite{Gross92}  

There is rather good agreement between four relativistic impulse
approximation calculations and the new $t_{20}$ data from Jefferson
Laboratory.  The relativistic calculations of Hummel and 
Tjon\cite{Hummel899094}, Van Orden, Devine and Gross\cite{VanOrden95}
and Phillips, Wallace and Devine\cite{PWD98} all show that the
$t_{20}$ data are well described by the impulse approximation.
Similarly good agreement has been obtained by Carbonnel and Karmanov using
the light-front formalism.\cite{CarbonellKarmanov99}  

For $A(Q),$ the new data extend to $Q^2 =
$6(GeV/c)$^2$. Relativistic calculations based on the
impulse approximation underpredict the $A(Q^2)$ data at large $Q$.
However, calculations of Van
Orden, Devine and Gross that include the $\rho \pi \gamma$
meson-exchange current contribution are in good agreement with the data.
The calculations include the $\rho\pi\gamma$ exchange current using
a soft form factor, $F_{\rho\pi\gamma}(Q^2)$, for the 
$\rho \pi \gamma$ vertex that is motivated by a
quark model.  Calculations of Hummel and Tjon used a
vector-meson dominance model of the $\rho\pi\gamma$ form factor that does
not fall as rapidly with increasing $Q^2$, which results in $A(Q^2)$ being too
large at high $Q^2$.  Thus, the softer $\rho \pi \gamma$ form factor
is favored by the $A(Q^2)$ data.

Three types of meson-exchange currents can play a role in the electron
deuteron scattering, namely, $\rho\pi\gamma$, $\omega\sigma\gamma$,
and $\omega\eta\gamma$ currents.  In the calculations of Hummel and
Tjon, Van Orden, Devine and Gross and Phillips, Wallace and Devine, inclusion
of the $\rho\pi\gamma$ current gives a rather small effect in $t_{20}$
that tends to improve the agreement with data slightly.  However, the
$\omega\sigma\gamma$ meson-exchange current that has been considered
by Hummel and Tjon is ruled out by the new $t_{20}$ data.  It predicts
a significant shift of the impulse approximation result towards higher
$Q$ that is not consistent with experiment.  Contributions from the
$\omega\eta\gamma$ current are rather small and they may be omitted.
Thus, the new $t_{20}$ data do not indicate much importance 
for meson-exchange contributions except at the highest $Q$.  
This is very interesting
because the meson-exchange currents, particularly the  
$\omega \sigma \gamma$ current,
contain substantial uncertainties.

Because the $\omega\sigma\gamma$ current is
uncertain, it has been omitted in many analyses.  It was introduced
by Hummel and Tjon in order to achieve a reasonable description of the
magnetic form factor, $B(Q)$, which is poorly described either by the
relativistic impulse approximation or when the $\rho\pi\gamma$
exchange current is included.  There is a surprising effect, first
pointed out by Zuilhof and Tjon\cite{Zuilhof79.81}, that causes relativistic calculations
of the magnetic form factor to deviate from the data.  When the
$\rho\pi\gamma$ current is included, the standard form of the
coupling of the $\rho$-meson to the nucleon, i.e., $g_{\rho
NN}(\gamma^\mu + \kappa/M\sigma^{\mu\nu}q_\nu)$, arises as a factor in
the $\rho\pi\gamma$ exchange current between two nucleons.  In a $p/M$
expansion, the tensor term proportional to $\kappa$ is higher order
and often is omitted.\cite{Riska89,ARW00}  However, the tensor term can be important
because $\kappa \approx 7$, making the tensor coupling of relative
order $Q/m_\pi$, rather than $Q/M$.  In relativistic calculations, the
tensor term generally is included.  The result is that there is a sign
change of the $\rho\pi\gamma$ exchange-current contribution to the
magnetic form factor near $Q^2 \approx $ 1 (GeV/c)$^2$.  This sign
change occurs at a lower $Q^2$ than the minimum of the magnetic form
factor, which is near $Q^2 \approx $ 2 (GeV/c)$^2$.  Because of the 
sign change, the $\rho\pi\gamma$ current shifts the minimum of $B(Q^2)$
to lower $Q$.
Without the sign change, the
$\rho\pi\gamma$ exchange current shifts the minimum of $B(Q^2)$ 
toward higher $Q$, which agrees better with the data.  This
effect has now been seen in a number of calculations.  Hummel and Tjon
incorporate the tensor term, with a consequent shift of $B(Q)$ away
from the data, and 
Phillips, Wallace and Devine\cite{PWD99} have reproduced the effect using the same
$\rho\pi\gamma$ current.  Recent calculations by Van Orden also show
the same effect when the $\rho\pi\gamma$ current is included.
Schiavilla\cite{SchiavillaPC} has independently checked this and has seen the same
effect.  

The sign change in the $\rho\pi\gamma$ contribution would not have an
adverse effect if it were to occur at higher $Q$, for example, $Q^2 > $2 (GeV/c)$^2$.  
A smaller ratio of $\rho NN$ tensor to vector coupling would
help to move the sign change to higher $Q^2$.

Because the new $t_{20}$ data seem to rule out a significant
contribution from the $\omega\sigma\gamma$ current, some other
contribution is required in order to explain the
magnetic form factor.  It has been found by Gross, Devine and Van
Orden that Z-graph contributions are large enough to provide
an explanation.  Including the negative-energy components of the
deuteron's wave function, as calculated in the spectator formalism,
provides a substantial shift of the calculated minimum of the $B(Q)$ 
form factor towards higher $Q$.  Attempts to confirm this result
by Phillips and Wallace and also by Tjon have not been successful.
The ET calculations show only rather small effects of negative-energy
components and no significant shift of the minimum of $B(Q^2)$.
This is surprising because the ET propagator 
provides a factor of two enhancement of
negative-energy components in comparison with the symmetrized
propagator of the spectator formalism, simply because 
the one-body limit of the ET propagator gives the Dirac propagator 
for each of the two nucleons.
Thus, negative-energy
components should be larger in ET calculations than in spectator
calculations, other things being equal.
Perturbative calculations have verified the small effects 
of Z-graphs in ET calculations.
Thus, there is no
convergence regarding the explanation of $B(Q)$ and further work on
the subject is warranted.

The deuteron form factor as defined by $F_{d}(Q^2) \equiv \sqrt{A(Q^2)}$
is observed to decrease as 
$Q^{-10}$ at large $Q$.\cite{JlabHallA-A(Q)}.  
This simple scaling behavior has been predicted from
perturbative QCD, but generally $Q$ is thought
to be too low for the perturbative QCD explanation to be
valid.\cite{Isgur80.89,Radyushkin84}  
Similar behavior can be obtained from 
a meson-exchange calculation as indicated by the successful description 
of $A(Q^2)$ by Van Orden and Gross.   

In summary the relativistic impulse approximation calculations provide
good agreement with the $t_{20}$ data up to $Q^2 \approx$ 1.5
(GeV/c)$^2$.  The data for $A(Q^2)$ up to $Q^2 \approx$ 6(GeV/c)$^2$
can be described when the $\rho\pi\gamma$ exchange current is included.
Results for $B(Q^2)$ form factor show much larger deviations from the
data, indicating that a new contribution may be needed in the
theoretical models.  There is not yet a convergence on what the new
contribution is.

\section{PHOTODISINTEGRATION OF THE DEUTERON}

\begin{figure}
\vbox to 2.5 truein {\vss
\hbox to 6.5 truein {\includegraphics{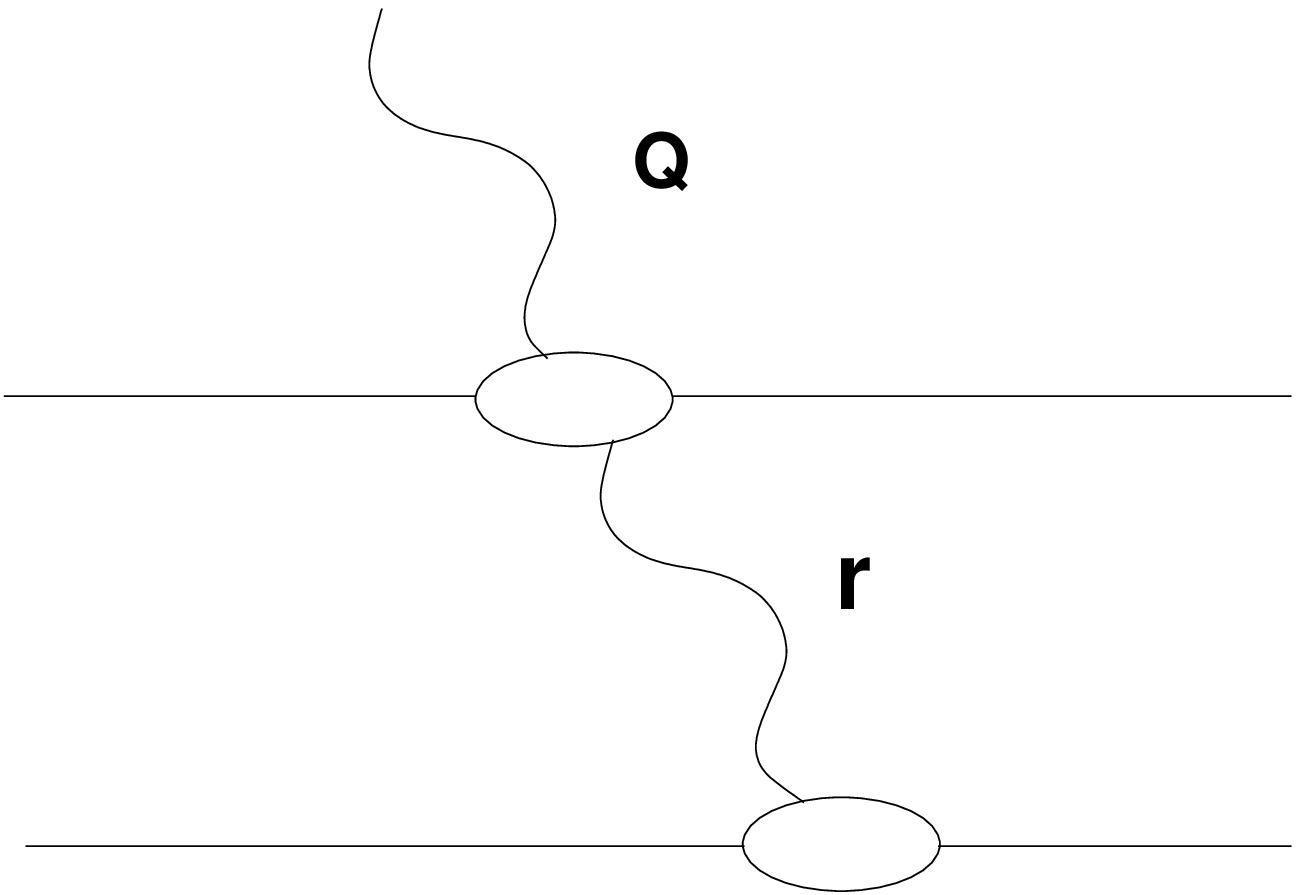}\hss}}
\caption{ Photon absorption followed by meson exchange between
the nucleons in a deuteron
\label{fig:Fig1}}
\vbox to 4.5 truein {\vss
\hbox to 6.5 truein {\includegraphics{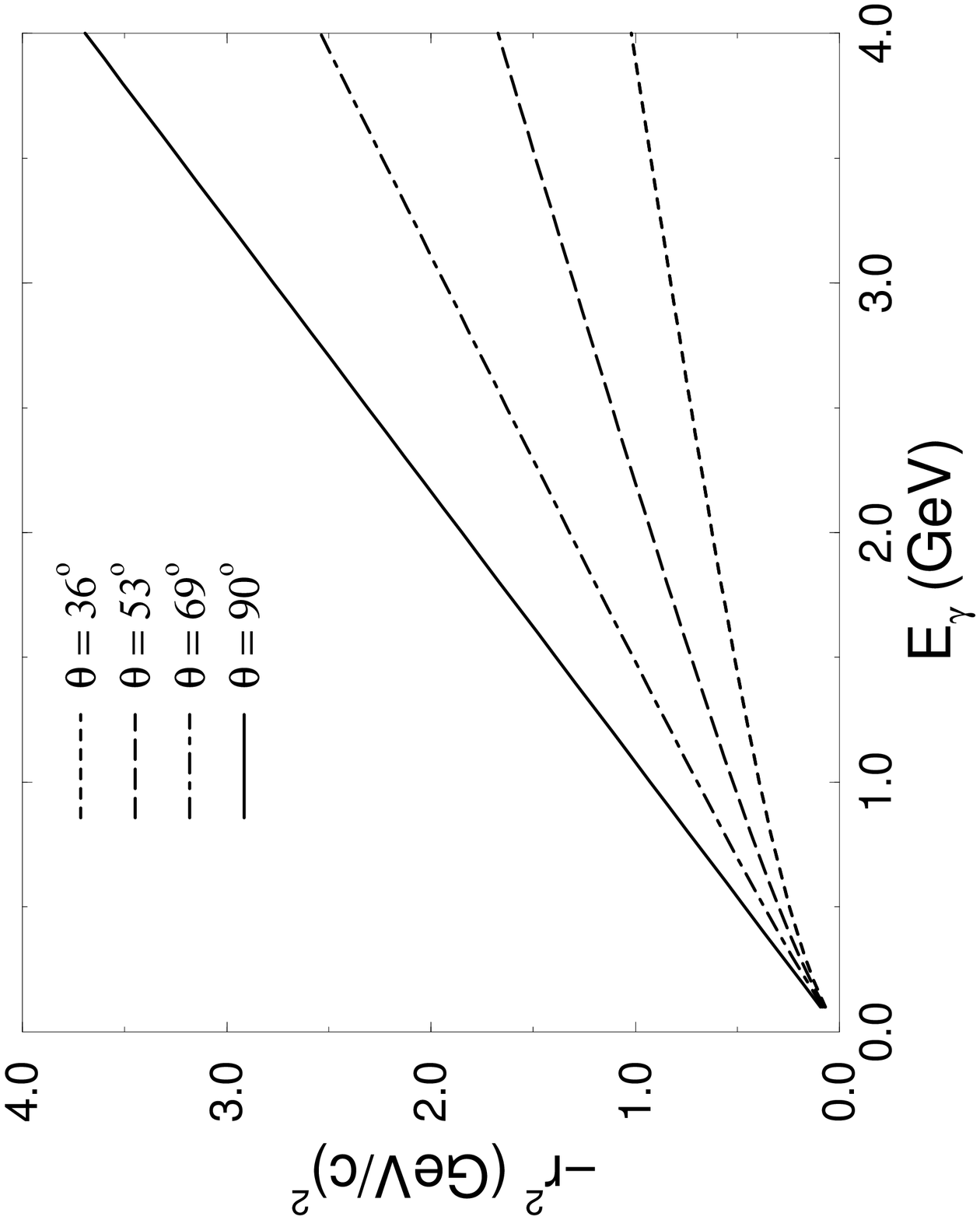}\hss}}
\caption{ Variation of momentum transfer with photon energy.
\label{fig:Fig2}}
\end{figure}

Very interesting experiments at Jefferson Laboratory have measured
cross sections and induced polarization in photodisintegration of the
deuteron with photons up to 4 GeV.\cite{Bochna98}  
The initial results were for
final state nucleons at $\theta_{c.m.} = 36^{\small o}, 52^{\small o},
69^{\small o}$ and $89^{\small o}$, where $\theta_{c.m.}$ is the angle 
between the final-state proton and neutron in their center of mass.    
At the largest angle, cross sections exhibited
scaling behavior as a function of the invariant mass squared $s = 4M^2
+ 2ME_\gamma$, i.e., $s^{11}d\sigma/dt \approx$ constant.  As the
angle between the nucleons was decreased, the cross sections deviated
from scaling behavior.  One way to understand this is in terms of the
momentum transfer $r$ in the diagram shown in Fig. 1 which depicts a
generic exchange mechanism between the two nucleons.  
Approximating
the two nucleons in the initial-state deuteron as each having zero
momentum leads to a simple expression for the momentum transfer,
i.e., $r^2 = -ME_{\gamma}(1-\alpha cos\theta_{c.m.})$,
where $\alpha = 1/\sqrt{1 + M/E_{\gamma}}$.  Thus, the momentum
transfer is spacelike and is largest in magnitude when 
$\theta_{c.m.} = 90^{\small o}$.  It decreases rapidly
as the angle becomes smaller as shown in Fig. 2.
At $\theta_{c.m.} = 36^{\small o}$,
the smallest angle in the experiment, momentum transfer is not greater
than about 1 (GeV/c)$^2$.   
For $E_{\gamma} = 4 GeV$ and $\theta_{c.m.} = 90^{\small o}$, 
$r^2 \approx -3.7 (GeV/c)^2$.  Thus the experiment spans a large
range of momentum transfer between the proton and neutron.

The most recent results show that the induced polarization of the
proton tends to zero when the photon energy is greater than $1 GeV$
at $\theta_{c.m.} = 90^{\small o}$.\cite{Wijesooriya00}  
Polarization transfer observables tend to zero for $E_{\gamma} > 2 GeV$.
Thus, at the largest momentum transfers in the experiment,
there is evidence that the process is consistent with
helicity conservation.  
A theoretical explanation of the cross section data
has been developed by Frankfurt, Miller, Sargsian and 
Strikman\cite{FMSS} by
relating the process to a hard $pn$ scattering process.  

The photodisintegration reaction probes interactions over a large
range of momentum transfers.  At lower momentum transfer, it is
expected to be dominated by hadronic interactions.  At the highest
momentum transfer, partonic interactions may be important in the subamplitude
where the photon is absorbed.  
A very interesting question is whether a transition occurs
from hadronic to partonic interactions  
within the range of photon
energies considered in the recent experiments.

\section{SUMMARY}

Very precise data have been obtained for the bremsstrahlung 
process in $pp$ collisions.  They are sufficient to explore
interesting effects of relativity and meson-exchange
currents and to determine whether the conventional meson-exchange
lagrangian is adequate.  Theoretical calculations based 
on dynamical models or on soft-photon approximations do not yet 
provide a consistent understanding of the new data.

Progress in threshold production of $\pi^0$ mesons in $pp$ collisions
has been made by measuring spin observables, which allow
a separation of s-wave and p-wave parts of the amplitude. At 
energies close to threshold where the s-wave contributions dominate,
experimental cross sections are about a factor of two larger
than the most complete theoretical cross sections, which are based on 
contributions of soft pions and intermediate $\Delta$-states.  
A short-range contribution is need to provide the missing 
strength.  The exact nature of the short range contributions is 
still debated, but it is consistent with being a mixture of relativistic 
effects and short-range meson-exchange currents.  

New data from Jefferson Laboratory have provided indications that
relativistic models based on meson-exchange forces work
surprisingly well to large $Q^2$.  Particularly the $t_{20}$ data 
show that short-range meson-exchange currents must play
a fairly small role up to $Q^2 \approx 1.5 (Gev/c)^2$.  
The $\rho\pi\gamma$ exchange current
is needed at large $Q^2 > 2 - 3 (Gev/c)^2$ to describe the $A(Q^2)$ data, but
the calculations based on the relativistic impulse approximation
are reasonable for $A(Q^2)$ and quite close to the 
experimental data for $t_{20}$.
However, there is a puzzle regarding the magnetic form factor.
The contribution
of the $\rho\pi\gamma$ exchange current that helps for
$A(Q^2)$ worsens the agreement between theory and experiment for
$B(Q^2)$.  
Relativistic effects could make a significant contribution, but the
existing analyses provide a conflicting assessment of their
role.   
Thus, a consistent picture has not yet been established.  

Intriguing experimental results have been obtained for
deuteron photodisintegration at photon energies up to $4~GeV$.
Cross sections exhibit scaling behavior at the largest momenta 
of the experiment, but these momenta are too low for perturbative
QCD to be valid.  Possibly a 
subamplitude of the process becomes dominated by partonic
interactions at the momentum transfers involved.  For example, 
the initial photon absorption by a nucleon might be dominated 
by a partonic amplitude, followed by a meson exchange to 
share the momentum transfer with the second nucleon.  
A variety of theoretical possibilities should be explored 
in order to understand the deuteron photodisintegration 
reaction and to investigate a possible transition 
from hadronic interactions to partonic interactions at high photon
energy.

\end{document}